\documentclass[manuscript]{aastex}
\usepackage[colorlinks,
            linkcolor=red,
            anchorcolor=blue,
            citecolor=blue
            ]{hyperref}

\shorttitle{Magnetic separatrix as the source region of an active-region filament}
\shortauthors{Zou, P. et al.}

\begin{document}

\title{Magnetic separatrix as the source region of the plasma supply for an active-region filament}

\author{P. Zou\altaffilmark{1,2,3}, C. Fang\altaffilmark{1,2,3},
 P. F. Chen\altaffilmark{1,2,3}, K. Yang\altaffilmark{1,2,3},
 \& Wenda Cao\altaffilmark{4}}

\altaffiltext{1}{School of Astronomy \& Space Science, Nanjing University, Nanjing 210023, China}
\altaffiltext{2}{Key Lab of Modern Astronomy \& Astrophysics (Nanjing University),
                Ministry of Education, Nanjing 210023, China}
\altaffiltext{3}{Collaborative Innovation Center of Modern Astronomy and Space Exploration,
                Nanjing 210023, China}
\altaffiltext{4}{Big Bear Solar Observatory, New Jersey Institute of Technology,
             ~40386 North Shore Lane, Big Bear City, CA 92314, USA}

\begin{abstract}
Solar filaments can be formed via chromospheric evaporation followed by
condensation in the corona or by the direct injection of cool plasma from the
chromosphere to the corona. In this paper, with high-resolution
H$\alpha$ data observed by the 1.6 m New Solar Telescope of the Big Bear Solar
Observatory on 2015 August 21, we confirmed that an active-region filament is
maintained by the continuous injection of cold chromospheric plasma. We find that
the filament is rooted along a bright ridge in H$\alpha$, which corresponds to
the intersection of a magnetic quasi-separatrix layer with the solar surface.
This bright ridge consists of many small patches and the sizes of these
patches are comparable to the width of the filament threads.
It is found that upflows originate from the brighter patches of the ridge,
whereas the downflows move toward the weaker patches of the ridge. The whole
filament is composed of two opposite directional streams, implying that
longitudinal oscillations are not the only cause of the counterstreamings, and
unidirectional siphon flows with alternative directions are another possibility.
\end{abstract}


\section{Introduction} \label{sec1}

Solar filaments are elongated dark structures appearing on the solar disk
typically observed in H$\alpha$ \citep{pare14, vial15}. They show similar, but 
more extended, dark features in extreme ultraviolet (EUV) lines or continuum
\citep{hein01, schm04}. When they are seen above the solar limb they are
called prominences, it is revealed that filaments are dense clouds suspended in
the corona. While their temperature is $\sim$100 times smaller than that of the
surrounding corona, their density is $\sim$100 times higher than the latter,
leading to the total mass of a filament in the range of $10^{14}$--$10^{15}$ g
\citep{labr10}. The mass is so high that the mass of several quiescent
filaments is comparable with the whole mass of the corona \citep{malh89}. In
other words, the mass of a filament cannot be supplied by the condensation of
the local corona itself \citep{mack10}. Therefore, the filament material must
originate from the chromosphere.

Several mechanisms have been proposed to explain how the chromosphere supplies
the necessary mass into the supporting magnetic structure in
the corona, either a flux rope or a sheared arcade.
One popular mechanism is the evaporation-condensation model \citep{pol86}, i.e.,
chromospheric plasmas are heated up to the coronal temperature and then
evaporated into the corona along the magnetic field line. As the hot plasma
accumulates, the coronal loop finally becomes so dense that the criterion for
thermal instability is satisfied, and the coronal plasma cools down to
$\sim$10$^4$ K in a catastrophic way \citep{park53}. This model can explain the
sudden appearance of many filaments, as illustrated by radiative hydrodynamic
or hydromagnetic simulations in one dimension \citep{kar01}, two dimensions
\citep{xia12}, and even three dimensions \citep{xia16}. The multiwavelength
observations in EUV confirmed the cooling process from $\sim$1.5 MK all the way
to 30,000 K in $\sim$8 hr \citep{liu12}. Alternatively, the cold chromospheric
material can be directly injected into the corona along the magnetic field lines,
as observed by \citet{wang01} and \citet{chae03}. The injection may result from
magnetic reconnection in the low atmosphere as magnetic elements with the
minority polarity approach the dominant polarity \citep{wang01}. This model is
also supported by accumulating observational evidence, which shows that the
endpoints of a filament or its barbs correspond to mixed magnetic polarities,
and cold jets flow toward the filament spine \citep{zou16}. It is noted,
however, that the mixed polarities are not the only place where magnetic
reconnection can happen. A more general case would be the magnetic
quasi-separatrix layers (QSLs), where current sheets can be formed easily and
reconnection can happen even when a single polarity is present on the
photospheric magnetogram. In this {\em Letter}, with the combined
high-resolution observations of the \emph{Solar Dynamics Observatory} ({\it
SDO}) and the 1.6 m New Solar Telescope \citep[NST,][]{cao10,goo12}, at Big
Bear Solar Observatory (BBSO), we report that the thread clusters of an
active-region filament are rooted along a bright ridge in H$\alpha$, which
corresponds to the intersection of a magnetic QSL with the solar surface. The
observations are described in \S\ref{obs}, and the results are presented in
\S\ref{res}, which are followed by discussions in \S\ref{dis}.

\section{Observations and Data Analysis} \label{obs}

\begin{figure}
\centering
\includegraphics[width=13cm]{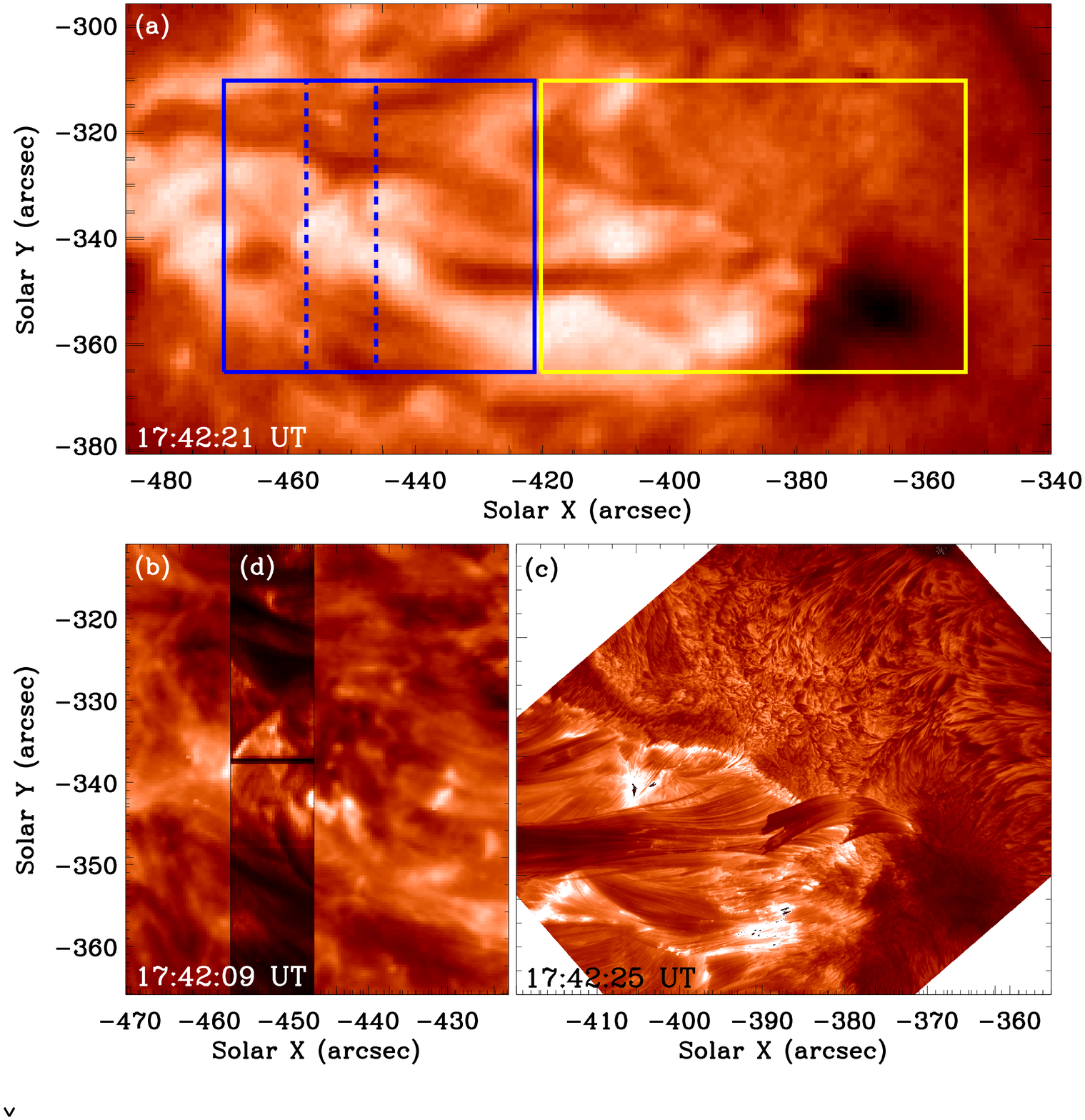}
\caption{The filament cannot be observed in H$\alpha$ images before our observation period (panel (a)).
Panel (b) gives the position of the filament. The FOV of each instrument is also shown. The FOV of NST
covers the west half of the filament (panel (d)) and that of the {\it IRIS} covers the other half (panel (c)).
The scanning region can observe one endpoint of the filament (panel (e)).}
\label{fig1}
\end{figure}

A small filament is located in the active region NOAA 12403 (S27E24) on 2015
August 21, as shown by Figure \ref{fig1}(a), which is the H$\alpha$ image from
the Global Oscillation Network Group \citep[GONG,][]{harv11}. This filament was
observed by the BBSO/NST from 17:00 UT to 19:00 UT. Its field of view (FOV)
covers only the western half of the filament, as indicated by the yellow box
in Figure \ref{fig1}(b). The {\it Interface Region Imaging Spectrograph}
\citep[{\it IRIS};][]{de14} mission observed the eastern half of the filament, with
both imaging in a wider FOV and spectroscopy in a smaller FOV as indicated by
the blue solid and dashed boxes in Figure \ref{fig1}(b).

For observing the fine structures of the filament, the Visible Image Spectrometer (VIS)
onboard NST, which uses a Fabry-P\'erot etalon with the bandpass of 0.07 \AA, is
required. It provides the H$\alpha$ line-center,  $\pm$0.2 \AA, $\pm$0.4 \AA,
$\pm$0.6 \AA, $\pm$0.8 \AA, and $\pm$1.0 \AA\ filtergrams. The FOV is about
$70\arcsec \times 70\arcsec$ and the image resolution is $0 \farcs 030$ per
pixel and the cadence is 32 s. The photospheric motions can be observed by
the Broad Filter Imager (BFI) which uses a TiO filter (7057 \AA) with a
bandpass of 10 \AA. Its spatial resolution is $0 \farcs 034$ per pixel and
the cadence is 15 s.

The \ion{Mg}{2} and \ion{Si}{4} spectra observed by {\it IRIS} are used for spectral
diagnostics of the filament. Its raster width is $0 \farcs 35$ and the spectral
resolution is 0.051 \AA\ per pixel for \ion{Mg}{2} line and 0.025 \AA\ per pixel
for \ion{Si}{4} respectively. {\it IRIS} provides the slit-jaw images of near ultraviolet
emission (2796 \AA\ and 2832 \AA) with the bandpass of 4 \AA\ and Far Ultraviolet
emission (1330 \AA\ and 1400 \AA) with the bandpass of 40 \AA. Both the slit-jaw
images have a pixel size of $0 \farcs 33$.

In order to understand the magnetic environment and the underlying mechanism of the filament
activity, we use the magnetograms observed by the Helioseismic and Magnetic Imager
\citep[HMI,][]{sch12,schou12} on board the \emph{Solar Dynamics Observatory} {\it SDO}.
The magnetograms have a resolution of $0 \farcs 60$ per pixel and a cadence of 45 s.
Furthermore, magnetic extrapolation can help us determine more detail about the
magnetic configuration. Data from the Space-weather HMI Active Region Patches
\citep[SHARPs,][]{bob14} are used. The SHARPs data have resolved the 180$^\circ$
ambiguity by adopting the minimum energy method \citep{met94,met06}. The coordinate
system has been modified by the Lambert method \citep{bob14}, and the project effects are
corrected with the method described by \citet{gar90}. In order to coalign the
observed data from different telescopes, both the BBSO/NST H$\alpha$ line wing
images and the slit-jaw images are compared with the continuum images observed
by {\it SDO}/HMI.

\section{Results}
\label{res}

Figure \ref{fig2}(a) displays a single snapshot of the H$\alpha$ images at line
center. It is clear that the filament is composed of many thin threads,
terminating at the western endpoints of the filament. The attached animation
shows that these threads are highly dynamic. New threads are formed continuously,
and then move along the direction of the threads. A clear impression is that the
upper part of the filament is dominated by rightward-moving threads, which drain
down to the solar surface, whereas the lower part of the filament is associated
with leftward-moving threads, which originate from the western endpoints of the
filament. In order to see the dynamics more clearly, we select two slices, i.e.,
yellow and blue slices, along the upper half and the lower half of the filament,
respectively, as marked in Figure \ref{fig2}(a). For both slices, the starting
points are chosen to be the western endpoint of the filament. The time evolution
of the intensity along the yellow and blue slices are plotted in panels (b) and (c) of
Figure \ref{fig2}, respectively. In order to show the flows clearly,
the image of the H$\alpha$ line wing is used. It is seen that
along the yellow slice, most of the intensity patterns show a motion toward the starting
point of the slice, as indicated by dashed lines in Figure \ref{fig2}(b). In
contrast, the blue slice is associated with sporadic ejections moving away from the
starting point, as indicated by dashed lines in Figure \ref{fig2}(c).
According to the time-distance diagrams, we find that these flows are 
sporadic, with an interval of $\sim$20 min. Each time, the flow lasts for 
about 10 min.

It is noted in Figure \ref{fig3}(a) that all of the threads are rooted along
a bright ridge in H$\alpha$. The ridge is not uniform in brightness, with some
parts fainter than others. Interestingly, the yellow slice corresponds to the threads
rooted at the fainter part of the ridge, whereas the blue slice corresponds to the
threads mapped to the brighter part. In order to understand the magnetic
properties of this bright ridge, we superpose the H$\alpha$ intensity contour
lines of its brightest kernels on the {\em SDO}/HMI magnetogram in Figure
\ref{fig3}(c). Apparently the bright ridge in H$\alpha$ extends from the eastern
edge of a positive sunspot all the way down to a region close to the magnetic
polarity inversion line (PIL) between a negative polarity and a positive polarity
further south. It is noted that the bright kernels have typical sizes
of $0 \farcs 1 \sim 0 \farcs 2$, which are comparable to the width of the filament
threads. These fine structures can only be seen in high-resolution observations.

With the {\em SDO}/HMI SHARPS vector magnetograms, we extrapolate
the nonlinear force-free coronal magnetic field by applying the optimization
method \citep{wie04}. Once the 3-dimensional coronal magnetic field is obtained,
its topological parameters, e.g., the squashing factor $Q$ \citep{tito02}, can
be calculated. As discussed by \citet{demo06}, the locations with extremely large
$Q$ correspond to QSLs, across which the magnetic connectivity changes
drastically. The distribution of $Q$ in the solar surface is plotted in Figure
\ref{fig3}(d). Comparing panels (a) and (d), it is seen that the bright ridge in
H$\alpha$ is associated with large $Q$, meaning that the H$\alpha$ bright ridge
corresponds to the footpoint of a QSL.

The left panel of Figure \ref{fig4} displays the {\it IRIS} slit-jaw image in
\ion{Mg}{2} k 2796 \AA\, which covers the eastern part of the filament. Because
of its wide bandpass, the filament is nearly invisible. Even though, it is seen
that the eastern endpoints of the filament are also rooted at an area with
continuing brightenings. A narrow region with $-457\arcsec\leq x\leq -448\arcsec$ is measured by
the {\it IRIS} spectrometer with the scanning mode. It takes $\sim$200 s to
complete one raster. Panels (b) and (c) in Figure \ref{fig4} display the
Dopplergrams at two times separated by $\sim$1 hr. It is seen that blue shifts
persist around the position (-456\arcsec, -345\arcsec) as indicated by the blue
arrows, whereas red shifts persist around the position (-448\arcsec,
-345\arcsec) as indicated by the red arrows. The blue shifts, i.e., upflows, have
typical velocities of $\sim$ 5 km s$^{-1}$, and the red shifts, i.e., downflows, have
typical velocities of $\sim$ 10 km s$^{-1}$.

The non-linear force free field (NLFFF) extrapolation of the filament magnetic
configuration is given in Figure \ref{fig5}. On comparing with the magnetic field
lines in Figure \ref{fig5} with Figure \ref{fig4}, it is revealed that the
blue-shifted area is roughly cospatial with the pink field lines which are
mapped to the yellow dashed line marked in Figure \ref{fig2}(a), whereas the
red-shifted area is associated with the yellow field lines which are
mapped to the blue dashed line in Figure \ref{fig2}(a). The Doppler shift
pattern is consistent with the extrapolated magnetic connectivity, i.e.,
cold upflows are ejected at the eastern footpoints of the pink field lines
in Figure \ref{fig5}, and then move westward until draining down to the
western footpoint of the pink field lines. In contrast, another stream of
cold upflows is ejected at the western footpoints of the yellow field lines,
and then move eastward until draining down to the eastern footpoints of the
yellow field lines.

\section{Discussions}\label{dis}

It is well known that filament material should originate from the
chromosphere. However, it has been debated whether the cold chromospheric
material is ejected into the corona directly as claimed by the injection
model \citep{chae03}, or is heated up to $\sim$2 MK and evaporated to the
corona where thermal instability leads to catastrophic cooling to
$\sim$10$^4$ K, as claimed by the evaporation-condensation model \citep{pol86}.
With the solid observations showing both direct injection \citep{zou16} and
in-situ formation via thermal instability \citep{liu12}, it seems that both
mechanisms are at work in the solar corona.

For the injection model, it has been proposed that the injected upflows are
cold jets resulting from the magnetic reconnection in the solar lower
atmosphere \citep{wang99, chae03}. One unclear issue regarding this model is how
reconnection can drive cool material to a height of $\sim$100 Mm in the
corona without heating the plasma too much \citep{mack10}. According to
\citet{jian12}, when the reconnection point is located in the low
chromosphere, most of the released magnetic energy goes to the kinetic energy
of the reconnection outflow, and the plasma is heated weakly. Furthermore,
radiation and ionization consume a significant part of the thermal energy
\citep{chen01}. These numerical simulations indicate that it is possible to
have fast but cold jets from the chromosphere to the corona. For the
evaporation-condensation model, however, the upper chromospheric plasma
should be heated. Although in most numerical models \citep{kar01, xia12, xia16},
a localized heating is specified in the chromosphere without mentioning which
mechanism, it is perceivable that the heating is also due to magnetic
reconnection. This means that both of the models for the filament formation
require magnetic reconnection in the chromosphere, and canceling magnetic
features have been discovered to be associated with the mass supply for
filaments \citep{wang01, chae03, zou16}. With this {\em Letter}, we emphasize that
the mixed polarities are not the only place which can host chromospheric
magnetic reconnection and inject cold plasma to feed a filament.
A more general scenario for the reconnection to happen is the magnetic QSL.
At least for the active-region filament observed on 2015 August 21, its endpoint is
found to be rooted at a magnetic QSL, which is calculated from the extrapolated
magnetic field and manifested by a bright ridge in H$\alpha$. Our
high-resolution observation reveals that the bright ridge consists of
many small kernels with sizes comparable to the width of the filament threads.
It implies that the mass supply of the filament threads may come from the
bright kernels.

Different formation mechanisms of solar filaments require different magnetic
configurations. For the filaments described by the evaporation-condensation
model, a magnetic dip is necessary for a flux tube, otherwise the condensed
plasma can not be suspended in the corona, and would drain down to the solar
surface along the field line upon formation. However, for the filaments formed
via continuous chromospheric injection, they are maintained dynamically with
quasi-continuous siphon flows as demonstrated by \citet{wang99} and the
observations in this paper. In this case, magnetic dips are not necessary.
They may or may not exist in the solar corona. It seems that this kind of
dynamically-sustained filaments may be of quiescent type \citep{wang99} or of
active-region type (this paper).

\begin{figure}
\centering
\includegraphics[width=15cm]{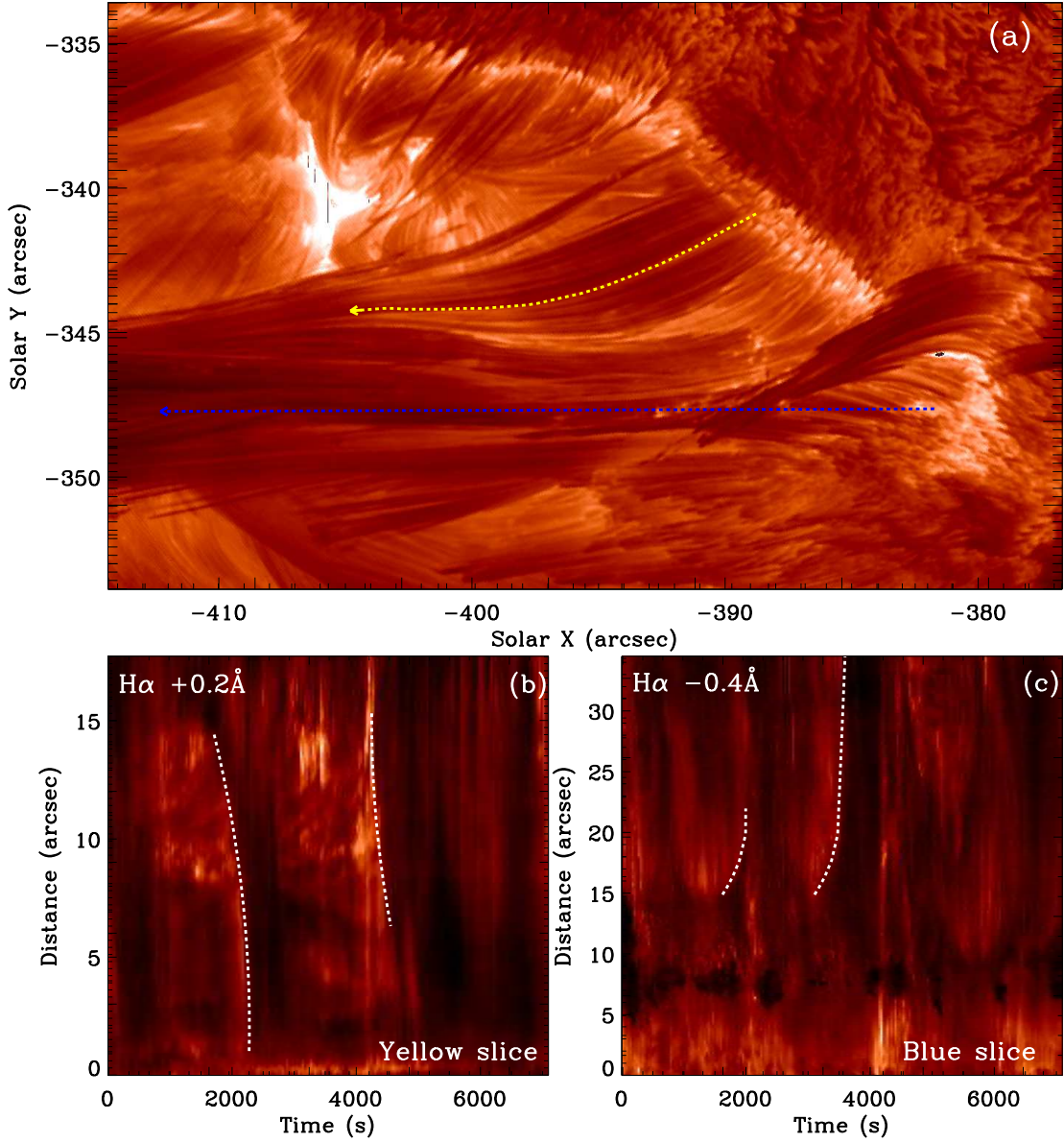}
\caption{Positions of the yellow and blue slices (top panel) and time-distance plots of the yellow
slice (panel (b)) and that of the blue slice (Panel (c)). The attached animation also shows the
dynamic motion of filament threads.}
\label{fig2}
\end{figure}

\begin{figure}
\centering
\includegraphics[width=10cm]{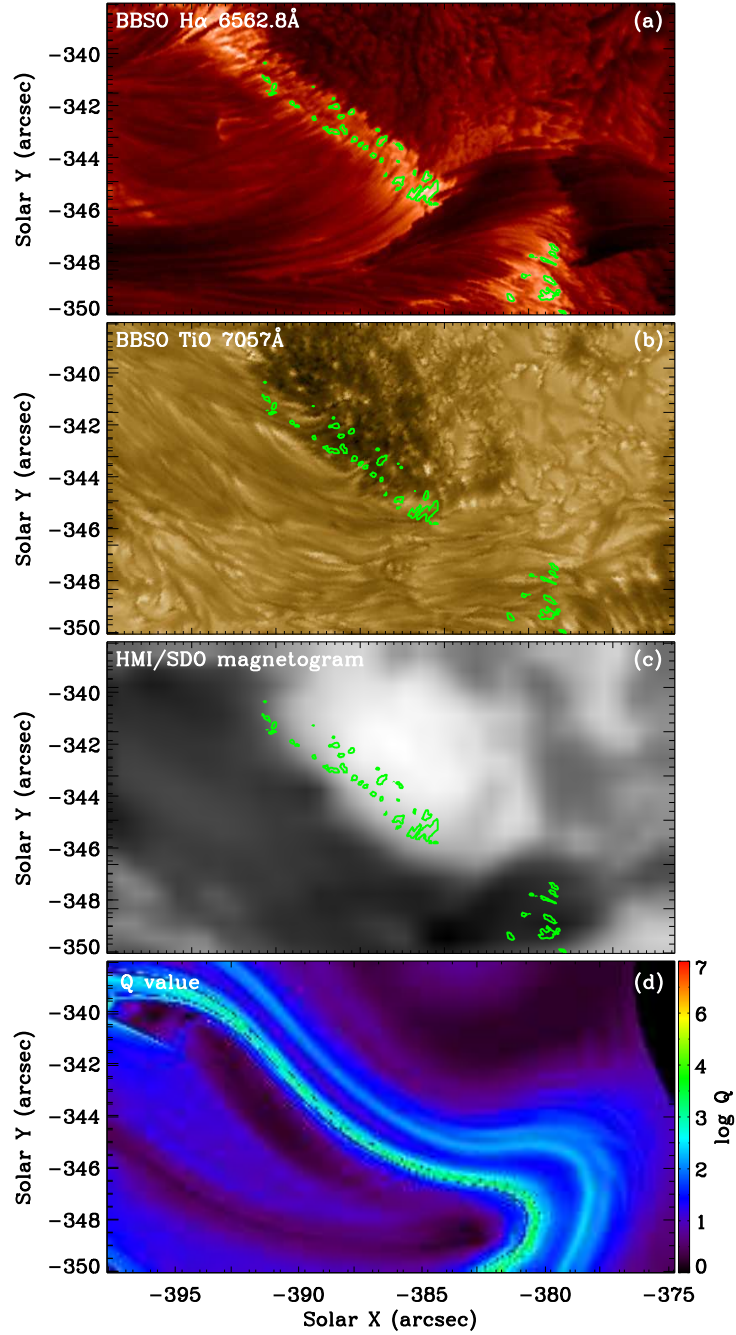}
\caption{Contours of bright patches on the H$\alpha$ line center image (panel (a)), the photospheric
TiO line image (panel (b)), and the magnetogram (panel (c)). The bottom panel shows the Q values
of this area.}
\label{fig3}
\end{figure}

\begin{figure}
\includegraphics[width=17cm]{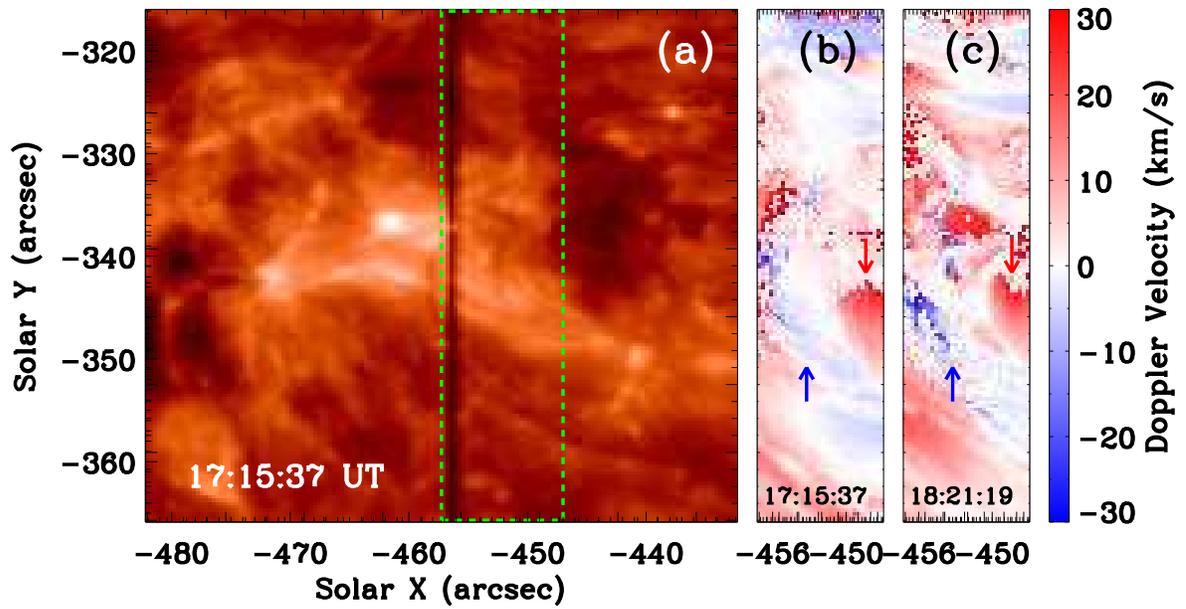}
\caption{Slit-jaw image of the \ion{Mg}{2} line and the scanning region (green-dashed box) (panel (a)).
The Dopplergrams of the scanning region at different times are shown in panels (b) and (c). The blue
and red arrows on each Dopplergram indicate the blue-shifted part and the red-shifted part respectively.}
\label{fig4}
\end{figure}

\begin{figure}
\includegraphics[width=17cm]{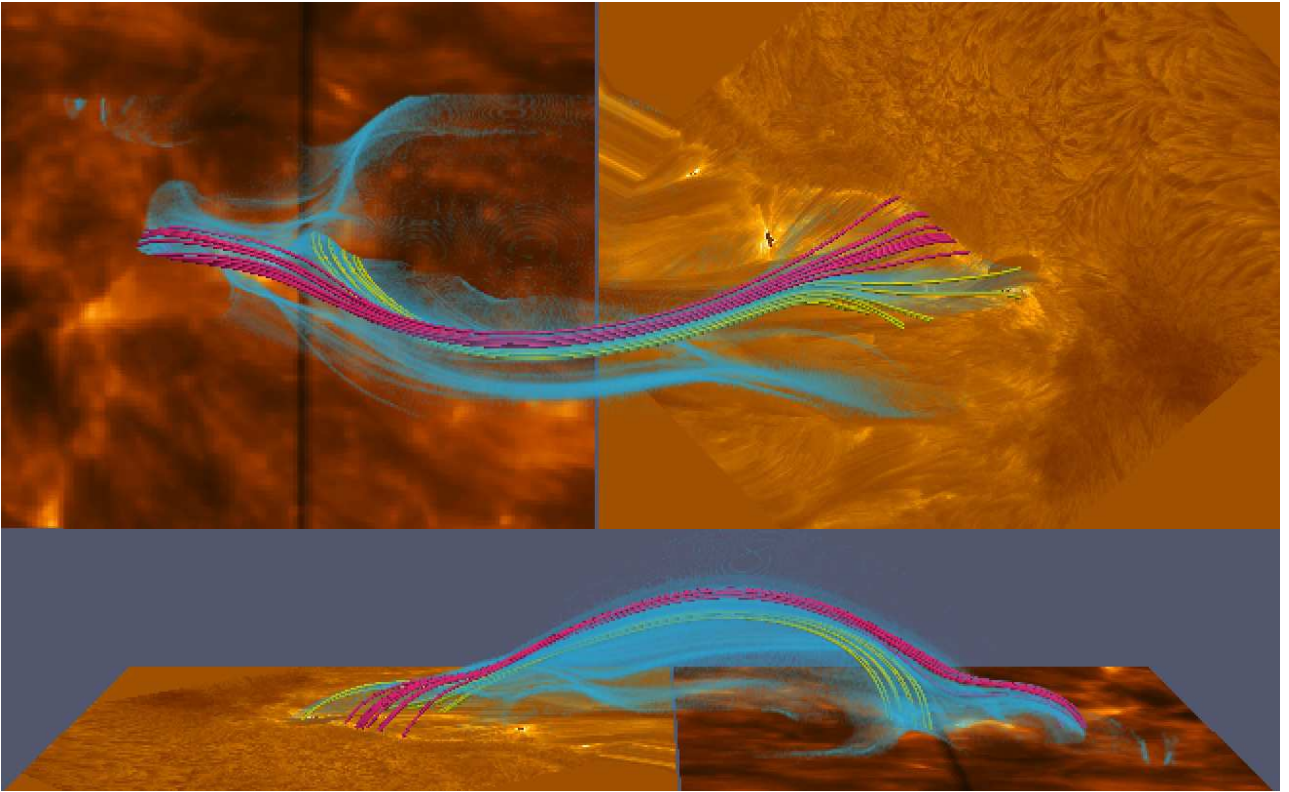}
\caption{Vertical (upper panel) and horizontal view (lower penal) of the filament magnetic configuration.
The pink and yellow magnetic lines correspond to different parts of the filament which have opposite
directional flows. The distribution of Q value is also shown by the cyan fog. The displayed Q values
are larger than 10$^{3}$.}
\label{fig5}
\end{figure}

A typical dynamic feature found in solar filaments is the counterstreaming
\citep{zir98}. Their nature has not yet been fully understood. In our
viewpoint, the nature of counterstreamings is not independent of the
formation mechanism and the magnetic configuration. In the case of
the evaporation-condensation model, the counterstreamings are probably
due to the longitudinal oscillation of the filament threads \citep{luna14},
where a magnetic dip exists near the top of the magnetic field line.
In the case of the chromospheric injection model, no longitudinal
oscillations are needed, nor are the magnetic dips. In this case, the
counterstreamings are the combination of unidirectional flows as implied
by the spectroscopic observations \citep{chen14}. In this paper, the
high-resolution H$\alpha$ observations clearly support such a scenario,
as revealed by Figure \ref{fig2}. It seems that the streaming is like a
siphon flow, moving from the brighter (hence with higher pressure)
footpoint to the weaker (hence with lower pressure) footpoint. As seen
from  Figure \ref{fig2}, the counterstreaming pattern of this filament
is very simple, with the upper half strand moving westward and
the lower half strand moving eastward.

\acknowledgments
The authors are grateful to the referee for the detailed suggestions. This 
work was supported by the National Natural Science Foundation of China
(NSFC) under the grant numbers 11533005, 11025314, and 13001003 as
well as NKBRSF under grant 2014CB744203. P.F.C. was also supported by Jiangsu
333 Project. W.C. acknowledges
the support of the US NSF (AGS-0847126 and AGS-1250818) and NASA (NNX13AG14G).
This work was also supported by the project ``The Strategic Priority Research
Program of the Chinese Academy of Sciences" (XDB09000000).

\clearpage

\begin{thebibliography}{}

\bibitem[Bobra et al.(2014)]{bob14} Bobra, M. G., Sun, X., Hoeksema, J. T., et al. 2014, Sol. Phys., 289, 3549
\bibitem[Cao et al.(2010)]{cao10} Cao, W., Gorceix, N., Coulter, R., et al. 2010, AN, 331, 636
\bibitem[Chae(2003)]{chae03} Chae, J. 2003, \apj, 584, 1084
\bibitem[Chen et al.(2001)]{chen01} Chen, P. F., Fang, C., \& Ding, M.-D., ChJAA, 1, 176
\bibitem[Chen et al.(2014)]{chen14} Chen, P. F., Harra, L. K., \& Fang, C. 2014, \apj, 784, 50
\bibitem[De Pontieu et al.(2014)]{de14} De Pontieu, B., Title, A. M., Lemen, J. R., et al.
        2014, Sol. Phys., 289, 2733D
\bibitem[Demoulin(2006)]{demo06} Demoulin, P., 2006, Advance in Space Research, 37, 1269
\bibitem[Gary \& Hagyard(1990)]{gar90} Gary, G. A., \& Hagyard, M. J. 1990, Sol. Phys., 126, 21
\bibitem[Goode \& Cao(2012)]{goo12} Goode, P. R., \& Cao, W. 2012, Proc. SPIE, 8444, 3
\bibitem[Harvey et al.(2011)]{harv11} Harvey, J. W., Bolding, J., Clark, R., et al., 2011,
        Bulletin of the American Astronomical Society, 1745
\bibitem[Heinzel et al.(2001)]{hein01} Heinzel, P., Schmieder, B., \& Tziotziou, K., 2001, \apj, 561, 223
\bibitem[Jiang et al.(2012)]{jian12} Jiang, R.-L., Fang, C., \& Chen, P. F., 2012, \apj, 751, 152
\bibitem[Karpen et al.(2001)]{kar01} Karpen, J. T., Antiochos, S. K., Hohensee, M., et al. 2001, \apj, 553, L85
\bibitem[Labrosse et al.(2010)]{labr10} Labrosse, N., Heinzel, P., Vial, J. C., et al. 2010,
         Space Science Review, 151, 243
\bibitem[Liu et al.(2012)]{liu12} Liu, W., Berger, T. E., \& Low, B. C., 2012, \apjl, 745, 21L
\bibitem[Luna et al.(2014)]{luna14} Luna, M., Knizhnik, K., Magluch, K, et al., 2014, \apj, 785, 79L
\bibitem[Mackay et al.(2010)]{mack10} Mackay, D. H., Karpen, J. T., Ballester, J. L., Schmieder, B.,
        \& Aulanier, G., 2010, Space Science Review, 151, 333
\bibitem[Malherbe(1989)]{malh89} Malherbe, J. M., 1989, In: Astrophysics and Space Science Library,
        Vol. 150, Dynamics and structure of quiescent solar prominences, ed. E. R. Priest, 115-141
\bibitem[Metcalf(1994)]{met94} Metcalf, T. R. 1994, Sol. Phys., 155, 235
\bibitem[Metcalf et al.(2006)]{met06} Metcalf, T. R., Leka, K. D., Barnes, G., et al. 2006, Sol. Phys., 235, 161
\bibitem[Parenti(2014)]{pare14} Parenti, S., 2014, LRSR, 11, 1
\bibitem[Park(1953)]{park53} Park, E. N., 1953, \apj, 117, 431
\bibitem[Poland \& Mariska(1986)]{pol86} Poland, A. I., \& Mariska, J. T., 1986, Sol. Phys., 104, 303
\bibitem[Scherrer et al.(2012)]{sch12} Scherrer, P. H., Schou, J., Bush, R. I., et al. 2012, Sol. Phys., 275, 207
\bibitem[Schmieder et al.(2004)]{schm04} Schmieder, B., Mein, N., Deng, Y., et al. 2004, Sol. Phys., 223, 119
\bibitem[Schou et al.(2012)]{schou12} Schou, J., Scherrer, P. H., Bush, R. I., et al. 2012, Sol. Phys., 275, 229
\bibitem[Titov et al.(2002)]{tito02} Titov, V. S., Horning, G. \& Demoulin, P., 2002, JGR, 107, 1164
\bibitem[Vial \& Engvold(2015)]{vial15} Vial, J.-C., \& Engvold, O. (ed.) 2015, 
	in Astro-physics and Space Science Library 415, Solar Prominences 
	(Berlin: Springer)
\bibitem[Wang(1999)]{wang99} Wang, Y.-M., 1999, \apj, 520, 71
\bibitem[Wang(2001)]{wang01} Wang, Y.-M., 2001, \apj, 560, 456
\bibitem[Wiegelmann(2004)]{wie04} Wiegelmann, T. 2004, Sol. Phys., 219, 87
\bibitem[Xia et al.(2012)]{xia12} Xia, C., Chen, P. F., \& Keppens, R., 2012, \apjl, 782, 26L
\bibitem[Xia \& Keppens(2016)]{xia16} Xia, C., \& Keppens, R., 2016, \apj, 823, 22
\bibitem[Zirker et al.(1998)]{zir98} Zirker, J. B., Engvold, O., \& Martin, S. F., 1998, Nature, 396, 440
\bibitem[Zou et al.(2016)]{zou16} Zou, P., Fang, C., Chen, P. F., et al. 2016, \apj, 831, 123

\end{thebibliography}
\end{document}